\documentclass[preprints,article,accept,moreauthors,pdftex]{Definitions/mdpi} 

\firstpage{1} 
\pubvolume{1}
\issuenum{1}
\articlenumber{0}
\pubyear{2021}
\copyrightyear{2020}
\datereceived{} 
\dateaccepted{} 
\datepublished{} 
\hreflink{https://doi.org/} 
\pdfoutput=1

\usepackage{amsmath}

\def\github{\url{https://github.com/anthony-walker/pysweep-git}}
\def\pysweep{\texttt{PySweep}}
\def\Swept{\texttt{Swept}}
\def\Standard{\texttt{Standard}}
\def\Up{\texttt{Up-Pyramid}}
\def\Down{\texttt{Down-Pyramid}}
\def\Oct{\texttt{Octahedron}}
\def\Xb{\texttt{X-Bridge}}
\def\Yb{\texttt{Y-Bridge}}

\def\oldCPU{Intel Skylake Silver 4114} 
\def\oldGPU{Nvidia GeForce GTX 1080 Ti}

\def\newCPU{Intel E5-2698v4} 
\def\newGPU{Nvidia Tesla V100-DGXS-32GB}

\Title{The Two-Dimensional Swept Rule Applied on Heterogeneous Architectures}

\TitleCitation{The Two-Dimensional Swept Rule Applied on Heterogeneous Architectures}

\Author{Anthony S.~Walker\orcidA{} and Kyle E.~Niemeyer\orcidB{}}

\AuthorNames{Anthony S.~Walker and Kyle E.~Niemeyer}

\AuthorCitation{Walker, A.S.; Niemeyer, K.E.}

\address[1]{%
School of Mechanical, Industrial, and Manufacturing Engineering\\
Oregon State University\\
Corvallis, Oregon, USA}

\corres{Correspondence: kyle.niemeyer@oregonstate.edu}

\abstract{The partial differential equations describing compressible fluid flows can be notoriously difficult to resolve on a pragmatic scale and often require the use of high performance computing systems and/or accelerators. 
However, these systems face scaling issues such as latency, the fixed cost of communicating information between devices in the system. The swept rule is a technique designed to minimize these costs by obtaining a solution to unsteady equations at as many possible spatial locations and times prior to communicating. 
In this study, we implemented and tested the swept rule for solving two-dimensional problems on heterogeneous computing systems across two distinct systems. Our solver showed a speedup range of 0.22--2.71 for the heat diffusion equation and 0.52--1.46 for the compressible Euler equations. 
We can conclude from this study that the swept rule offers both potential for speedups and slowdowns and that care should be taken when designing such a solver to maximize benefits. These results can help make decisions to maximize these benefits and inform designs.}


\begin{document}

\section{Introduction}
Partial differential equations (PDEs) are used to model many important phenomena in science and engineering. Among these, fluid mechanics and heat transfer are which can be notoriously difficult to solve on pragmatic scales. Wildfire is a good example of an expensive numerical simulation because it involves fluid mechanics, heat transfer, and chemical reactions over massive areas of impact. Wildfire is also an unsteady phenomenon---so it changes over time. The ability to accurately model and predict the properties of such phenomena in real time would allow for better responses, but there are many challenges associated with making these predictions.

Unsteady multi-dimensional PDEs often require using of distributed-memory computing systems to obtain a solution with practical grid resolution or scale in a reasonable time frame. Advancing the solution at any point in the grid inherently depends on the neighboring points in each spatial dimension. This dependence requires communication between computing nodes, to transfer data associated with the boundary locations. Each of these communications incurs a minimum cost regardless of the amount of information communicated---this is network latency. In contrast, bandwidth is the variable cost associated with the amount of data transferred. The total latency cost can significantly restrict the solution's time to completion, especially when using distributed systems. This barrier to scaling is referred to as the ``latency barrier'' and it impacts large-scale simulations that involve advancing many time steps, i.e., ones that require a large amount of communication \cite{Alhubail2016ThePDEs}. 
The barrier is a bottleneck in the system that can limit the performance regardless of the architecture.

Graphics processing units (GPUs) are powerful tools for scientific computing because they are well suited for parallel applications and large quantities of simultaneous calculations. Modern computing clusters often have GPUs in addition to CPUs because of their potential to accelerate simulations and, similar to CPUs, they perform best when communication is minimized. These modern clusters are often referred to as heterogeneous systems or systems with multiple processor types---CPUs and GPUs, in this case \cite{owens2007survey,owens2008gpu}. Ideally, a heterogeneous application will minimize communication between the GPU and CPU which effectively minimizes latency costs. Minimizing latency in high performance computing is one of the barriers to exascale computing which requires the implementation of novel techniques to improve \cite{Alexandrov2016RouteSkills}.

This study presents our implementation and testing of a two-dimensional heterogeneous solver for unsteady partial differential equations that employs a technique to help overcome the latency barrier---the swept rule \cite{Alhubail2016ThePDEs}. In other words, the swept rule is a latency reduction technique that focuses on obtaining a solution to unsteady PDEs at as many possible locations and times prior to communicating with other computing nodes (ranks). In this article, we first discuss related work, distinguish our work from prior swept studies, and provide more motivation in section~\ref{related-section}. Next, we describe implementation details, objectives, study parameters, design decisions, methods, and tests in section~\ref{methods-section}. As expected, this is followed with results and conclusions in sections~\ref{results-section} and~\ref{conclusions-section}.

\section{Related Work}
\label{related-section}
Surmounting the latency barrier has been approached in many ways; the most closely related to this study include prior swept rule studies which involve multiple dimensions and architectures but not the combination of the two \cite{Alhubail2016ThePDEs,Alhubail2018ThePDEs,Magee2018AcceleratingDecomposition,Magee2020ApplyingSystems}. Parallel-in-time methods are also related to the swept rule but differ in the sense that they iteratively parallelize the temporal direction, while the swept rule minimizes communication \cite{Gander201550Integration}. There are many examples of parallel-in-time methods which are all variations of the core concept \cite{Falgout2014ParallelMultigrid,Maday2020AnAlgorithm,Wu2018Parareal,EmmettTowardEquations,MinionINTERWEAVINGMULTIGRID,Hahne2020PyMGRIT:MGRIT}. For example, Parareal is one of the more popular parallel-in-time methods; it works by iterating over a series of coarse and fine grids with initial guesses to parallelize the problem \cite{Maday2020AnAlgorithm}. 
However, there are some stability concerns with Parareal that are addressed by local time integrators \cite{Wu2018Parareal}. These methods have the same goal but achieve that goal differently. Similarly, cache optimization techniques have the same objective, but they achieve it differently by optimizing communication not avoiding it \cite{Kowarschik2003AnAlgorithms}.

Finally, communication avoidance techniques closely emulate the swept rule but involve overlapping parts or redundant operations. The GPU implementation particularly blurs this difference because it solves an extra block to avoid intra-node communication but no extra blocks are solved for inter-node communication. The swept rule also differs because it particularly focuses on solving PDEs. There are varying degrees of communication algorithms that tend to focus on linear algebra \cite{DemmelAvoidingComputations, Ballard2011MinimizingAlgebra,BaboulinAMachines,Khabou2012LUVersion,SolomonikAHoefler}. 

The swept rule was originally developed by Alhubail et al.~\cite{Alhubail2016ThePDEs}, who presented a one-dimensional CPU-only swept PDE solver which was tested by solving the Kuramoto--Sivashinsky equation and the compressible Euler equations. 
They concluded that in each case a number of integrations can be performed during the latency time of a communication. Their analysis showed that integration can be made faster by latency reduction and increasing computing power of the computing nodes \cite{Alhubail2016ThePDEs}. 
Alhubail et al.~\cite{Alhubail2018ThePDEs} followed this work with a two-dimensional CPU only swept PDE solver which reported speedups of up to three times compared to classical methods when solving the wave and Euler equations. 
These studies differ from our study most prominently by the dimensionality and intended architecture.

Magee et al.\ created a one-dimensional GPU swept solver and a one-dimensional heterogeneous solver and applied both to solving the compressible Euler equations \cite{Magee2018AcceleratingDecomposition,Magee2020ApplyingSystems}. 
They concluded that their shared memory approach typically performed better than alternative approaches, but speedup was not obtained in all cases for either study. Varying performance results were attributed to greater usage of lower-level memory, which limits the performance benefits of the swept rule depending on the problem \cite{Magee2018AcceleratingDecomposition}.
Our current study extends upon the swept rule for heterogeneous architectures, but it differs in the dimensionality. Our implementation also attempts to use and extend some of the implementation strategies that showed promise in the aforementioned studies.

The effect of added dimensionality on performance is a pragmatic interest and can be considered from multiple perspectives. The primary goal is speeding up simulations requiring high performance computing by reducing network latency. The swept rule is motivated by reducing the time to obtain solutions of problems involving complicated phenomena frequently requiring the use of high performance computing systems. While many simplifications exist to reduce the dimensionality of fluid dynamics problems, most realistic problems are three-dimensional. Our solver is a step towards more realistic simulations by considering two spatial dimensions, which can provide insight into multi-dimensional performance and constraints. This insight can offer the chance to optimize system usage and promote faster design and prototype of thermal fluid systems.

In the event that computation time is not the primary concern, available resources or resource costs are important considerations. The ability to execute a simulation on a high performance computing system depends on access to such systems. 
Pay-per-use systems like Amazon AWS, Microsoft Azure, and Google Cloud offer cloud computing time. However, pricing models for university-managed clusters remain ambiguous making determination of cost challenging on the user's end \cite{Mesnard2019}. 
In the case of Amazon EC2, simulation time can be purchased at different hourly rates depending on the application. We generated an estimate using Amazon's pricing tool with a two node configuration (\texttt{g4dn.xlarge} instances) which yielded a monthly ``on-demand\\ cost of \$928.46 \cite{AmazonServices}. 
Purchasing such time and configures quickly becomes expensive for applications that require large numbers of computing hours or larger hardware configurations. Network latency contributes substantially to this cost as it is aggrandized in applications requiring a lot of communication because each communication event takes a finite amount of time regardless of the data size. 

Furthermore, it is possible to obtain and show performance benefits on smaller systems. This claim is supported by findings from Magee et al., who showed speedup on a workstation with a single GPU and CPU~\cite{Magee2018AcceleratingDecomposition}. 
While this is not the primary focus, an optimized solver that reduces latency would require less computing resources and more practical applications could potentially be solved on smaller, less costly computing clusters. 
Hopefully, it is clear at this point that latency reduction is important in high performance computing and scientific applications as this is the intention of this work.

\section{Materials and Methods}
\label{methods-section}

\subsection{Implementation \& Objectives}

We call our implementation of the two-dimensional swept rule \pysweep{}\footnote{\pysweep~is openly available at \github}. It consists of two core solvers: \Swept{} and \Standard{}. \Swept{} minimizes communication during the simulation via the swept rule. \Standard{} is a traditional solver that communicates as is necessary to complete a timestep, and serves as a baseline to the swept rule. Both solver use the same decomposition, process handling, and work allocation code so that a performance comparison between them is representative of swept rule performance. However, \Swept{} does require additional calculations prior to solving which are penalties of this swept rule implementation. 

We implemented \pysweep{} using Python and CUDA; the parallelism relies primarily on \texttt{mpi4py} \cite{DalcinMPIPython} and \texttt{pycuda} \cite{KlocknerPyCUDAGeneration}. Each process spawned by MPI is capable of managing a GPU and a CPU process, e.g., 20 processes can handle up to 20 GPUs and 20 CPU processes. Consequently, the aforementioned implementation allowed us to meet the objectives of this study on the swept rule, which include understanding:
\begin{enumerate}
    \item its performance on distributed heterogeneous computing systems,
    \item its performance with simple and complex numerical problems on heterogeneous systems,
    \item the impact of different computing hardware on its performance, and
    \item the impact of input parameters on its performance.
\end{enumerate}

\subsection{Parameters \& Testing}
\label{parameters-section}

GPUs execute code on a ``block-wise'' basis, i.e., they solve all the points of a given three-dimensional block simultaneously. We refer to the dimensions of these blocks as block size or $b$, which is a single integer that represents the $x$ and $y$ dimensions. 
The $z$ dimension of the block was always unity because the solver is two-dimensional. The block size is a parameter of interest because it affects the performance of the swept rule by limiting the number of steps between communications. 
It also provides a natural basis for decomposing the data---using multiples of the block size makes splitting the data among nodes and processors convenient because the GPU already requires it.

The swept solution process restricts the block size, $b$, to the range $(2n,\,b_{\max}]$ where $b_{\max}$ is the maximum block size allowed by the hardware and $n$ is the maximum number of points on either side of any point $j$ used to calculate the derivatives. We then define the maximum number of time steps that can be taken before communication is necessary as $k$:
\begin{equation}
    \label{blocksize-equation}
    k = \frac{b_{\max}}{2n}-1 \;.
\end{equation}
Consequently, we restrict the block size to being square ($x=y$) because the block's shortest dimension limits the number of time steps before communication. 

Blocks provide a natural unit for describing the amount of work, e.g., a $16\times16$ array has four $8\times8$ blocks to solve. As such, the work load of each GPU and CPU was determined by the GPU share, $s$, on a block-wise basis:  
\begin{equation}
    \label{share-equation}
    s = \frac{G}{W} = 1-\frac{C}{W} \;.
\end{equation}
where $G$, $C$, and $W$ represent the number of GPU blocks, CPU blocks, and total blocks, respectively. 
A share of 1 corresponds to the GPU handling 100\% of the work.
This parameter is of interest because it directly impacts the performance of the solvers.

In our implementation, the share does not account for the number of GPU or CPU cores available but simply divides the given input array based on the number of blocks in the $x$ direction, e.g., if the array contains $10$ blocks and $s=0.5$ then five blocks will be allocated as GPU work and the remainder as CPU work. These portions of work would then be divided amongst available resources of each type. 

Array size is another parameter of interest because it demonstrates how performance scales with problem size. We restricted array size to be evenly divisible by the block size for ease of decomposition. Square arrays were also only considered so array size is represented by a single number of points in the $x$ and $y$ directions. Finally, array sizes were chosen as multiples of the largest block size; this can result in unemployed processes if there are not enough blocks. Potential for unemployed processes is, however, consistent across solvers and still provides a valuable comparison.

The numerical scheme used to solve a problem directly affects the performance of the swept rule as it limits the number of steps that can be taken in time before communication. As such, the aforementioned parameters were applied to solving the unsteady heat diffusion equation and the unsteady compressible Euler equations as was done in prior swept-rule studies \cite{Alhubail2016ThePDEs,Alhubail2018ThePDEs,Magee2018AcceleratingDecomposition, Magee2020ApplyingSystems}. The heat diffusion equation was applied to a temperature distribution on a flat plate and solved using the forward-Euler method in time and a three point finite difference in each spatial direction. The compressible Euler equations were applied to the isentropic Euler vortex problem and solved using a second-order Runge-Kutta in time and a five-point finite volume method in each spatial direction with a minmod flux limiter and Roe-approximate Riemann solver \cite{SpiegelAMethods,Leveque2002FiniteProblems}. Appendices~\ref{Heat-Diffusion} and~\ref{Compressible-Euler} provide further detail on these methods and associated problems.

Hardware was selected based on the available resources at Oregon State University to analyze performance differences of the swept rule based on differing hardware. The methods were tested separately with 32 processes. The first two nodes each have \newGPU{} GPUs and \newCPU{} CPUs, and the second two nodes each have \oldGPU{} GPUs and \oldCPU{} CPUs. As a convention, parameters with respect to the first set of hardware will be referred to with a subscript $1$ and likewise a subscript $2$ for the second set. Each of these sets was used to solve both the heat diffusion equation and the compressible Euler equations. Section~\ref{results-section} present the performance data we collected. 

The results of this study raised some questions about the scalability of the algorithms used. Accordingly, weak scalability was considered with up to four nodes. Each node was given approximately two million spatial points and solved both problems for five hundred time steps. A share of 90\% and block size of 16 were used in this test. 

\subsection{Swept Solution Process}
\label{swept-process-section}

To recap, the swept rule is a latency reduction technique that focuses on obtaining a solution to unsteady PDEs at as many possible locations and times prior to communicating. The algorithm works by solving as many spatial points of a time step as it can and repeating this for subsequent time steps until information is required from neighboring nodes or hardware. This leaves the solution in a state with several incomplete time steps. Communication is then necessary to continue and fill in the incomplete time steps. This process repeats in various forms until the entire solution space has been completed to a desired time level. This process becomes more convoluted as we dive deeper into the details. To aid understanding, we refer to the two primary actions of the code as communication and calculation. We refer to the specific type of calculation as the phase. 

We considered communication to be when the code was transferring data to other ranks. Node-based rank communication between the CPU and GPUs happens via the use of shared memory, a capability implemented in MPICH-3 or later \cite{Hoefler2013MPIMemory}. Inter-node communication was performed with typical MPI constructs such as send and receive. A shared GPU memory approach implemented by Magee et al. showed some benefit \cite{Magee2018AcceleratingDecomposition}; this concept led to the idea of using shared memory on each node for node communication and primary processes for inter-node communication.
 
We considered calculation to be specifically when the code was solving in time. The specifics of CPU and GPU calculation depended on the particular phase the simulation was in. In general, the GPU calculation proceeded by copying the allocated section of the shared array to GPU memory and launched the appropriate kernel; the nature of GPU computing handled the rest. The CPU calculation began by disseminating predetermined blocks amongst the available processes. Each block was a portion of the shared memory which each process is allowed to modify.
 
There are four phases that occur during the swept solution process. In the code, we refer to them as \Up{}, \texttt{Bridge} (\texttt{X} or \texttt{Y}), \Oct{}, and \Down{}---that convention is continued here. We named them this way because these are the shapes produced during the solution procedure if you visualize the solution's progression in three dimensions $(x,y,t)$; this is demonstrated in later figures for a general case. To summarize the progression of swept phases, the \Up{} is calculated a single time; the \texttt{Bridge} and \Oct{} phases are then repeated until the simulation reaches a value greater than or equal to that of the desired final time. Finally, the \Down{} executes to fill in the remaining portion of the solution. 

Now that we have presented the major working components of the swept algorithm, we can dive into specifics of each phase and how they are defined. The specific shape that forms during each phase is a function of the spatial stencil, time scheme chosen to solve a given problem, and block size. As previously determined in section~\ref{parameters-section}, the maximum number of steps based on block size is $k$. A communication therefore must be made after $k$ steps for each phase with the exception of \Oct{} which communicates after $2k$ steps.
 
\begin{figure}[htbp]
    \centering
    \includegraphics[height=9cm,width=0.78\textwidth, trim={1cm 0.6cm 0.25cm 0.2cm},clip]{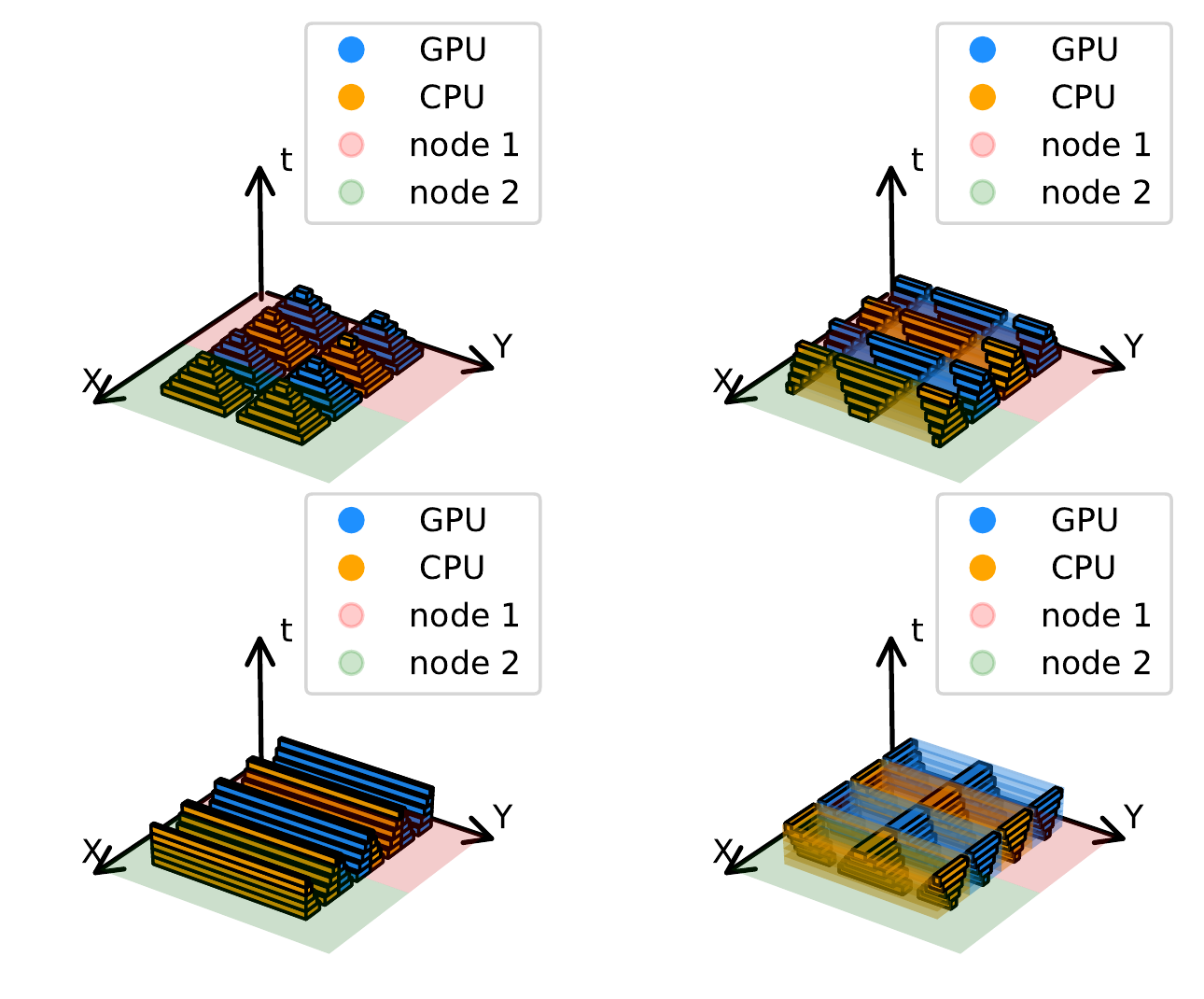}
    \caption{The first four steps in the swept solution process: \Up{} (top left), \Yb{} (top right), \texttt{Communication} (bottom left), and \Xb{} (bottom right).}
    \label{fig:MainOne}
\end{figure}

The first phase of calculation is the \Up{} shown in Figure~\ref{fig:MainOne} (top left). A dynamic programming approach was used on the CPU portion to calculate the \Up{} in lieu of conditional statements. Specifically, the indices to develop the \Up{} were stored in a set which was accessed as needed. The GPU portion implemented a conditional statement to accomplish this same task because its speed typically dwarfed that of the CPU regardless. If the code were to eventually be optimized, it could be beneficial to consider a dynamic programming approach such as that implemented on the CPU. In both cases, conditional or dynamic, the code removed $2n$ points from each boundary after every time step. If the time scheme required multiple intermediate time steps, these points were removed after each intermediate step. The initial boundaries of the \Up{} were the given block size and the final boundaries were found as $2n$ using a variation of equation~\ref{blocksize-equation}.

The next phase in the process is referred to as the Bridge which occurs independently in each dimension. The solution procedure for the bridges on the CPU and GPU follows the same paradigm as the \Up{} but twice as many bridges are formed because there is one in each direction. Bridges formed in each direction are labeled as such, e.g., the Bridge that forms parallel to the $y$ axis as time progresses is referred to as the \Yb{}. The initial boundaries of each Bridge were determined from $n$ and $b$. The Bridges grew by $2n$ from these initial boundaries in their reference dimensions and shrank by $2n$ in the other dimension which allowed the appropriate calculation points to be determined conditionally or prior to calculation.

Depending on the decomposition strategy, a large portion of the Bridges can occur prior to inter-node communication. In this implementation, the \Yb{}---shown in Figure \ref{fig:MainOne} (top right)---occurs prior to the first communication. The first inter-node communication then follows by shifting nodal data $b/2$ points in the positive $x$ direction. Any data that exceeds the boundaries of its respective shared array is communicated to the appropriate adjacent node. This process is demonstrated in Figure~\ref{fig:MainOne} (bottom left). The shift in data allows previously determined sets and blocks to be used in the upcoming calculation phases so the \Xb{} proceeds directly after the communication as shown in Figure~\ref{fig:MainOne} (bottom right).

\begin{figure}[htbp]
    \centering
    \includegraphics[height=9cm,width=0.78\textwidth, trim={1cm 0.6cm 0.25cm 0cm},clip]{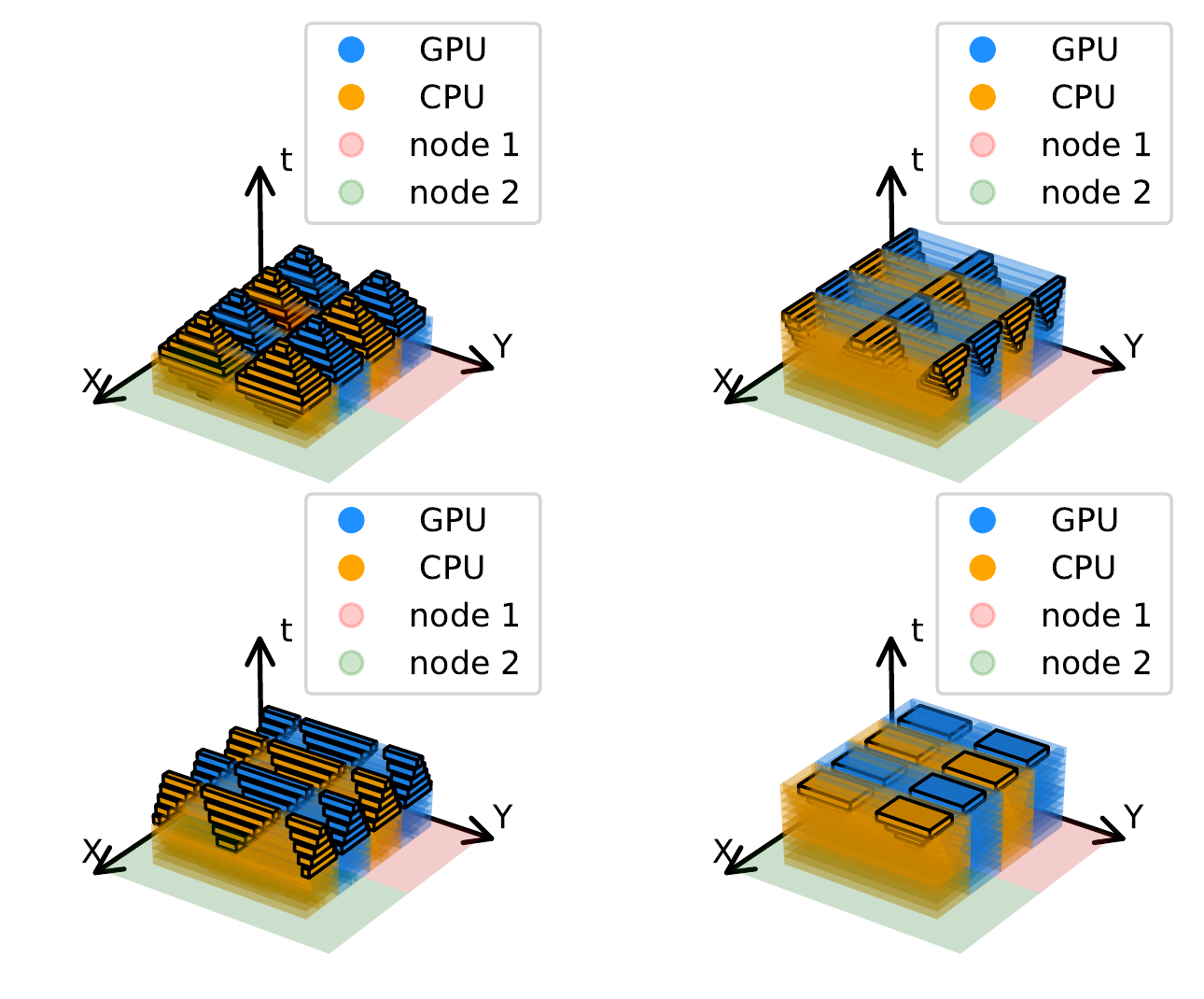}
    \caption{The intermediate and final steps of the swept solution: \Oct{} (top left), \Xb{} (top right), \Yb (bottom left), and \Down{} (bottom right).}
    \label{fig:MainTwo}
\end{figure}

The first three phases in the swept solution process demonstrated in Figure~\ref{fig:MainOne} are followed by the \Oct{} phase shown in Figure~\ref{fig:MainTwo} (top left). This phase is a superposition of the \Down{} and \Up{}. The \Down{}---shown in Figure~\ref{fig:MainTwo} (bottom right)--begins with boundaries that are $2n$ wide and expand by $2n$ on each boundary with every passing time step. This start is a natural consequence of removing these points during the \Up{} phase. The \Down{} completes upon reaching the top of the previous \Up{} or \Oct{} when the upward portion of \Oct{} phase begins. The entire \Oct{} is calculated in the same fashion as the \Up{} on both CPUs and GPUs. While the steps are described separately for clarity, they occur in a single calculation step without communication between ranks. 

The \Oct{} always precedes the \Yb{}, Communicate, \Xb{} sequence. However, the communication varies in direction as the shift and communication of data is always the opposite of the former communication. We repeated this series of events as many times as was necessary to reach the final desired time of each simulation. The final phase is the aforementioned \Down{} which occurs only once at the end of the simulation. We show the ending sequence---minus the communication---in its entirety in Figure~\ref{fig:MainTwo}. 


The final number of time steps taken by a swept simulation is determined by the number of \Oct{} phases ($2k$ time steps) that most accurately capture the specified number of time steps. This is a consequence of the swept rule: the exact number of steps is not always achievable in some cases because the simulation only stops after the completion of a phase. These phases occur on both the GPU and CPU with respect to the given share. Between each calculation step, a communication step occurs that consists of shared memory data management and writing to disk.

The shared-memory data management of the communication step as well as the writing to disk involve a couple of nuances worth mentioning. It includes shifting the data, which is a strategy implemented for handling boundary blocks in the array. \pysweep{} was implemented with periodic boundary conditions based on the targeted test problems. The boundary blocks of the array form half of the shape it would normally form in the direction orthogonal to the boundary (e.g., during the \Oct{} phase on the boundary where $x=0$, only half the \Oct{} will be formed in the $x$ direction). As expected, the corner will form a fourth of the respective shape. In lieu of added logic for handling these varying shapes, we implemented a data shifting strategy that allows the majority of the same functions and kernels to be used. The boundary points are able to be solved as if they were in the center of a block with this strategy. This strategy comes at the expense of moving the data in memory. 

\pysweep{} writes to disk during every communication as it is the ideal time. The code uses parallel HDF5 (h5py) so that each rank can write its data to disk independently of other ranks \cite{Collette2008HDF5Python}. The size of the shared memory array is $2k$ spaces in time plus the number of intermediate steps of the time scheme so that the intermediate steps of the scheme may be used for future steps if necessary. The appropriate fully solved steps are written to disk. The data is then moved down in the time dimension of the array so that the next phase can be calculated in the existing space.

The solver has a few restrictions based on architecture and implementation which have been previously described. It is currently implemented for periodic boundary conditions but can be modified to suit other conditions using the same strategy. The solver is also capable of handling given CPU functions and GPU kernels so that it may be used for any desired application that falls within the guidelines presented here. In this condition, we found it suitable for this study.

\section{Results}
\label{results-section}

Here we present the results of applying \pysweep{} to the heat transfer and compressible flow problems introduced in section~\ref{methods-section}. When solving these problems, we varied the following parameters to assess performance: array size, block size, share, and hardware. Array sizes of [320, 480, 640, 800, 960, 1120] were used to make the total number of points span a couple orders of magnitude. Block sizes of [8, 12, 16, 24, 32] were used based on hardware constraints. Share was varied from 0 to 100\% at intervals of 10\%. Along with these parameters, we advanced each problem 500 time steps.

\subsection{Heat Diffusion Equation}
\label{hdeResults}

We solved the heat diffusion equation numerically using forward-Euler and a three-point central difference in space. We verified it against an analytical solution developed for the two-dimensional heat diffusion equation with periodic boundary conditions---the problem formulation described in Appendix~\ref{Heat-Diffusion}. Figure~\ref{fig:heatSurface} compares the temperature distribution of the numerical and analytical solutions over time. The temperature distribution is representative of a flat plate with a checkerboard pattern of hot and cold spots. We see these spots diffuse out over time and approach a steady intermediate temperature which is expected. We also see that the numerical solution matches the analytical solution which verifies that we are solving the problem correctly.

\begin{figure}[htbp]
    \centering
    \includegraphics[height=9cm,width=0.78\textwidth, trim={0.9cm 0.3cm 0.1cm 0.9cm},clip]{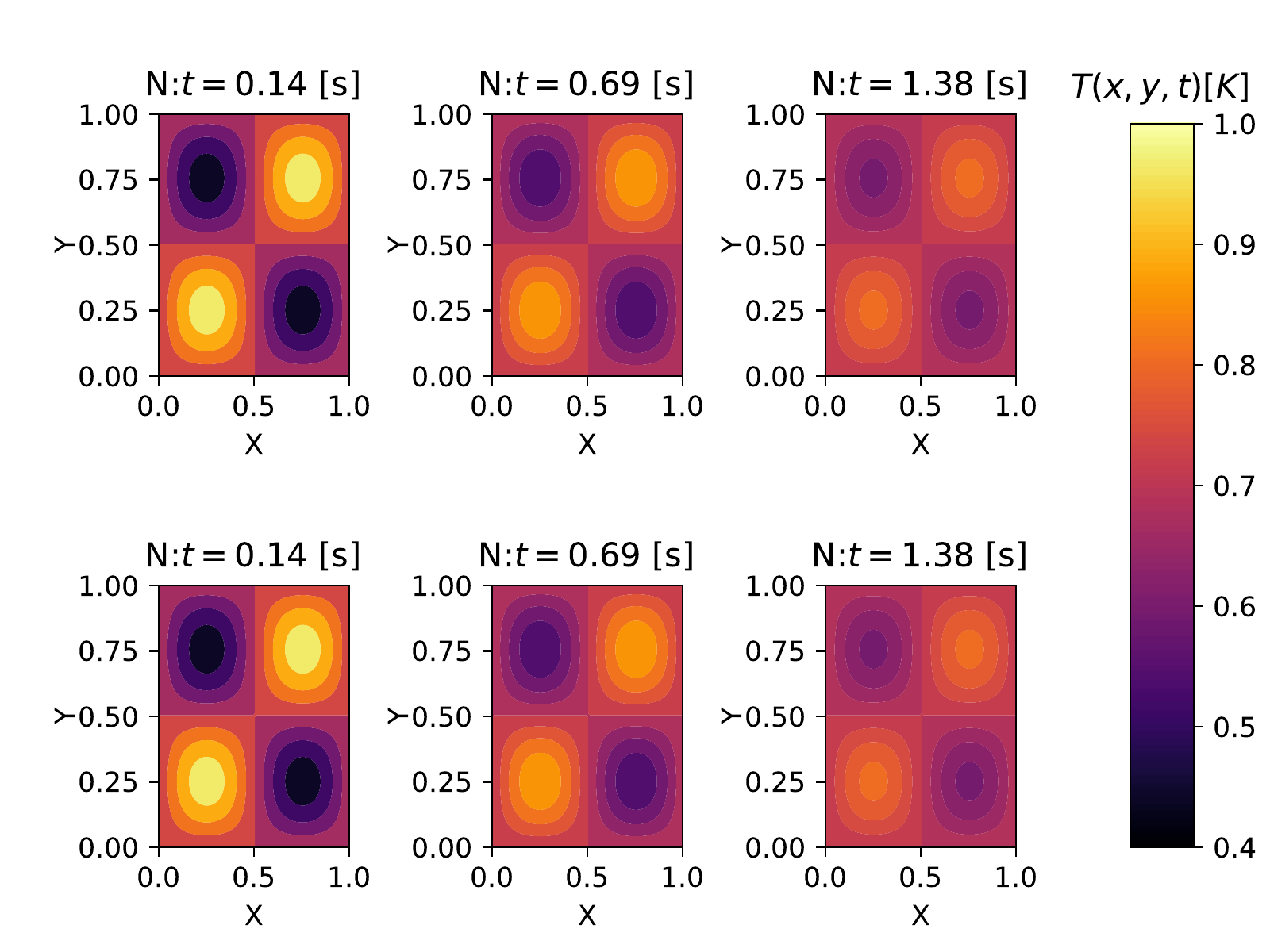}
    \caption{Verification of heat diffusion equation.}
    \label{fig:heatSurface}
\end{figure}

\begin{figure}[htbp]
    \centering
    \includegraphics[height=9cm,width=0.78\textwidth, trim={0.75cm 0.4cm 0.8cm 0.7cm},clip]{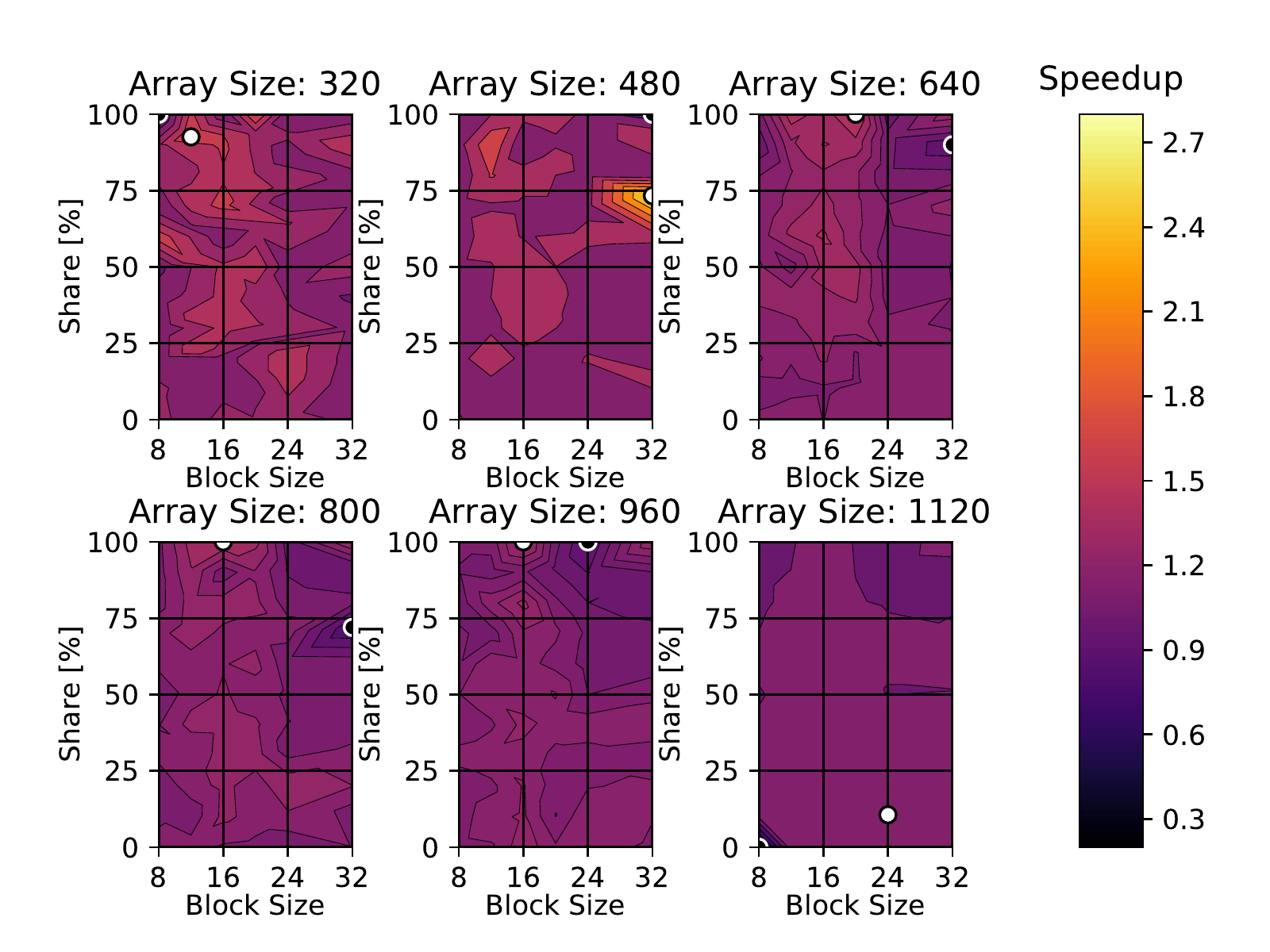}
    \caption{Swept rule speedup results for the Heat Diffusion Equation with \newGPU{} GPUs and \newCPU{} CPUs.}
    \label{fig:newSpeedupHeat}
\end{figure}

\begin{figure}[htbp]
    \centering
    \includegraphics[height=9cm,width=0.78\textwidth, trim={0.75cm 0.4cm 0.8cm 0.7cm},clip]{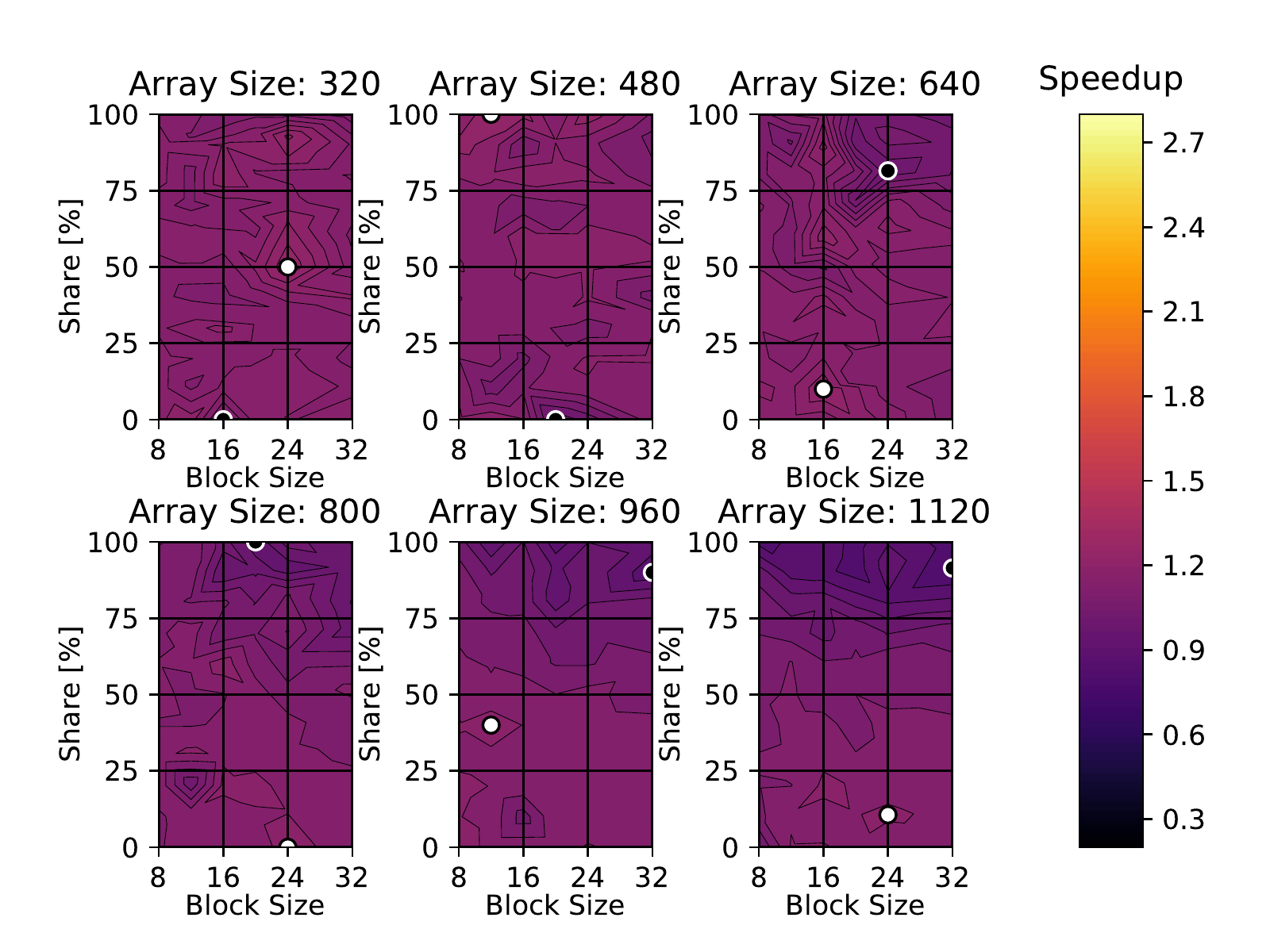}
    \caption{Swept rule speedup results for the Heat Diffusion Equation with \oldGPU{} GPUs and \oldCPU{} CPUs.}
    \label{fig:oldSpeedupHeat} 
\end{figure}

Our performance metric of choice for the swept rule was speedup, $S$, as a function of array size, block size, and share:
\begin{equation}
    S = \frac{R_{\textit{standard}}}{R_{\textit{swept}}} \;,
\end{equation}
where $R_i$ is the run time of simulation $i$.
Figures~\ref{fig:newSpeedupHeat} and~\ref{fig:oldSpeedupHeat} show the performance results produced from our tests for the two sets of hardware described in section~\ref{parameters-section}. 
In these figures, a black dot with a white border represents the worst case and a white dot with a black border represents the best case.

Figure~\ref{fig:newSpeedupHeat} shows a range of 0.22--2.71. 
It is clear that the performance of the swept rule diminishes as the array size increases. The largest areas of performance increases generally exist above a share of 50\% and along the block size of 12--20. The majority of best cases lie between 90--100\% share and the 12--20 block size range. 
The worst-performing cases lay close to the optimal cases: between 90--100\% share but outside the block size limits 12-20.

In Figure~\ref{fig:oldSpeedupHeat}, we see different trends than in Figure~\ref{fig:newSpeedupHeat} with a range of 0.79-1.32. 
Performance again diminishes as array size increases, it is better with lower shares in the case of this hardware, and it somewhat consistent across different block sizes. The best case for a given array size tends to occur at or below a 50\% share and between block sizes 12--24. 
The worst case for a given array size tends to occur between 80--100\% share and between block sizes of 20--32 with a couple occurrences on the upper limit.

\subsection{Compressible Euler Equations}
\label{eulerVortexResults}

\begin{figure}[htbp]
    \centering
    \includegraphics[height=9cm,width=0.78\textwidth, trim={0.75cm 0.3cm 0.2cm 0.2cm},clip]{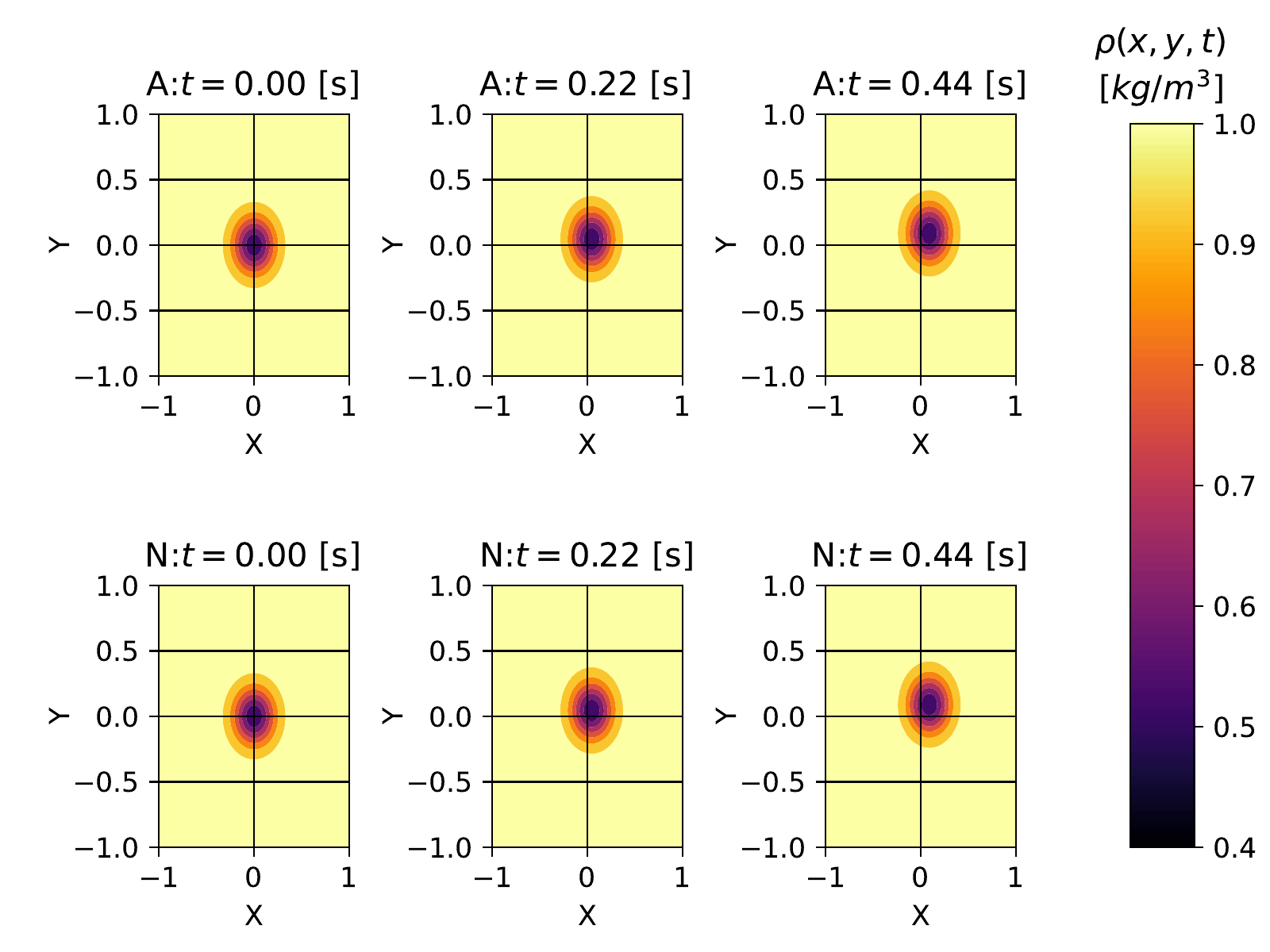}
    \caption{Verification of compressible Euler equations.}
    \label{fig:eulerSurface}
\end{figure}

\begin{figure}[htbp]
    \centering
    \includegraphics[height=9cm,width=0.78\textwidth, trim={0.75cm 0.4cm 0.8cm 0.7cm},clip]{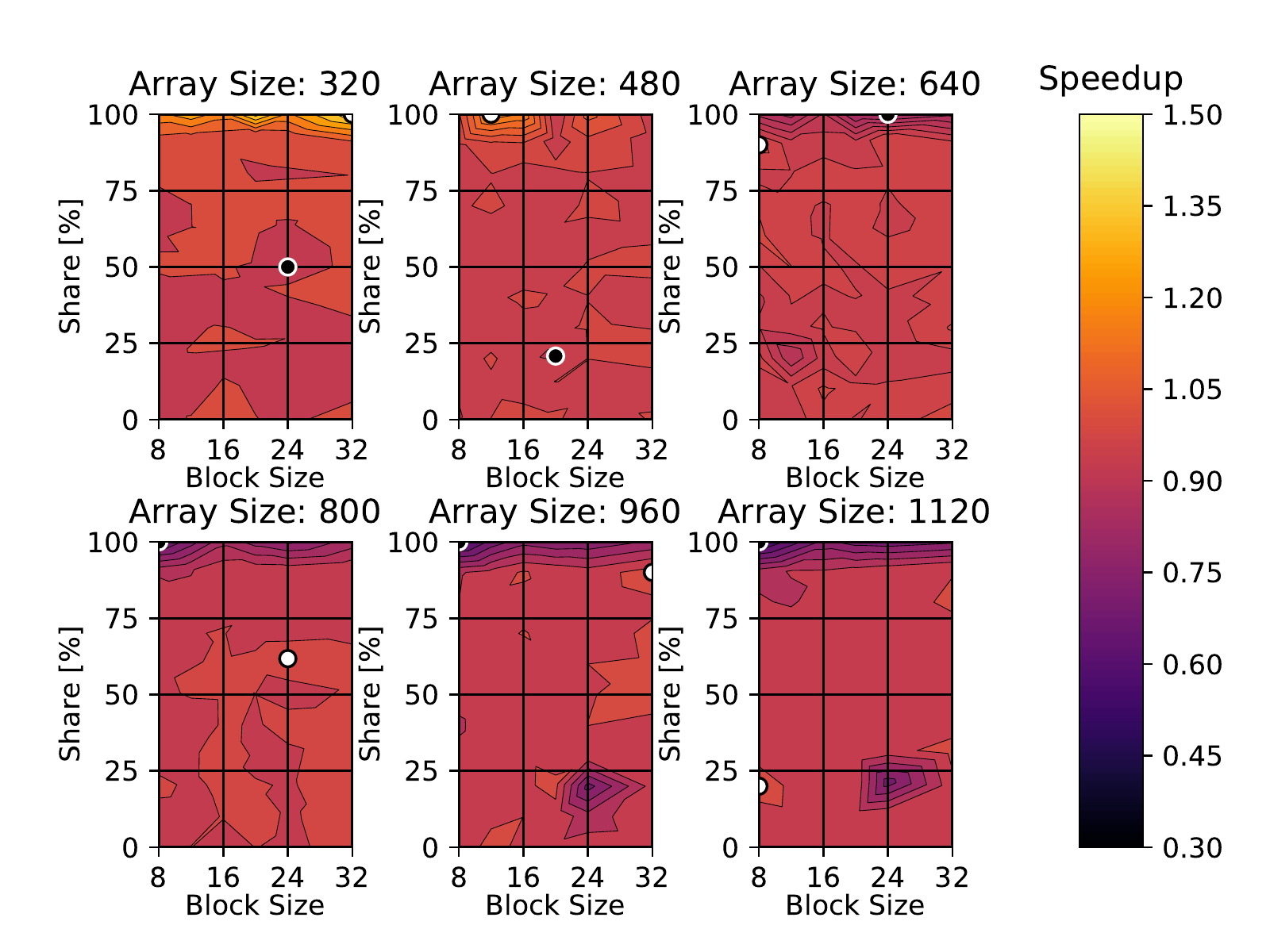}
    \caption{Swept rule speedup results  for the compressible Euler equations with \newGPU{} GPUs and \newCPU{} CPUs.}
    \label{fig:newSpeedupEuler}
\end{figure}

\begin{figure}[htbp]
    \centering
    \includegraphics[height=9cm,width=0.78\textwidth, trim={0.75cm 0.4cm 0.8cm 0.7cm},clip]{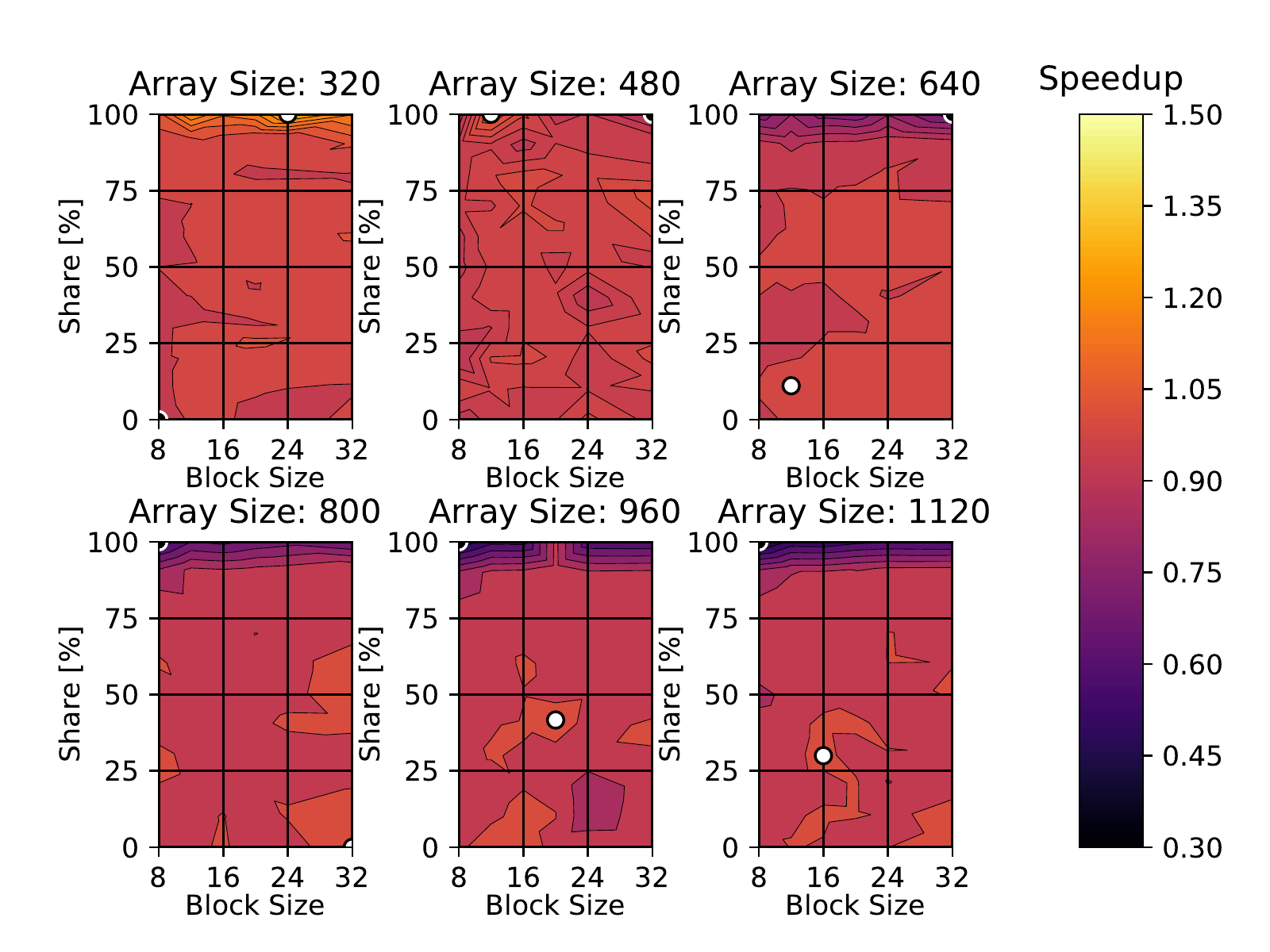}
    \caption{Swept rule speedup results  for the compressible Euler equations with \oldGPU{} GPUs and \oldCPU{} CPUs.}
    \label{fig:oldSpeedupEuler}
\end{figure}

We solved the Euler equations using a second-order Runge--Kutta scheme in time and a five-point central difference in space with a minmod flux limiter and Roe-approximate Riemann solver, which we verified against the analytical solution to the isentropic Euler Vortex described in Appendix~\ref{Compressible-Euler}. 
We performed the same tests here as in section~\ref{hdeResults}. Figures~\ref{fig:newSpeedupEuler} and~\ref{fig:oldSpeedupEuler} show performance results produced by these simulations. Figure~\ref{fig:eulerSurface} compares the numerical and analytical solutions of the density of the vortex over time. As we expect, the initial conditions of the vortex translate with an angle of $45^\circ$. This specific angle comes from the imposed velocity field. We also see that the solutions match.

Figure~\ref{fig:newSpeedupEuler} shows that the \Swept{} solver consistently performs slower than \Standard{} with a range of 0.52-1.46. Similar to the heat equation, performance declines with increasing array size but there is benefit in some cases. The best case for a given array size tends to occur above approximately 90\% share but there is no clear block size trend. The worst case for a given array size tends to occur at 100\% share but likewise a block size trend is unclear. However, in the three largest array sizes they always occur at 100\% share with a block size of 8.

Figure~\ref{fig:oldSpeedupEuler} shows that the \Swept{} solver is consistently slower than \Standard{} with a range of 0.36-1.42. Similar to the other configurations, performance consistently declines with increasing array size. Most best cases occur below approximately 50\% share with a block size between 12-24. The majority of the worst cases occur at 100\% share on the block size limits of 8 and 32.

\subsection{Weak scalability}

\begin{figure}[htbp]
    \centering
    \includegraphics[height=9cm,width=0.78\textwidth, trim={0cm 0.75cm 1cm 1.25cm},clip]{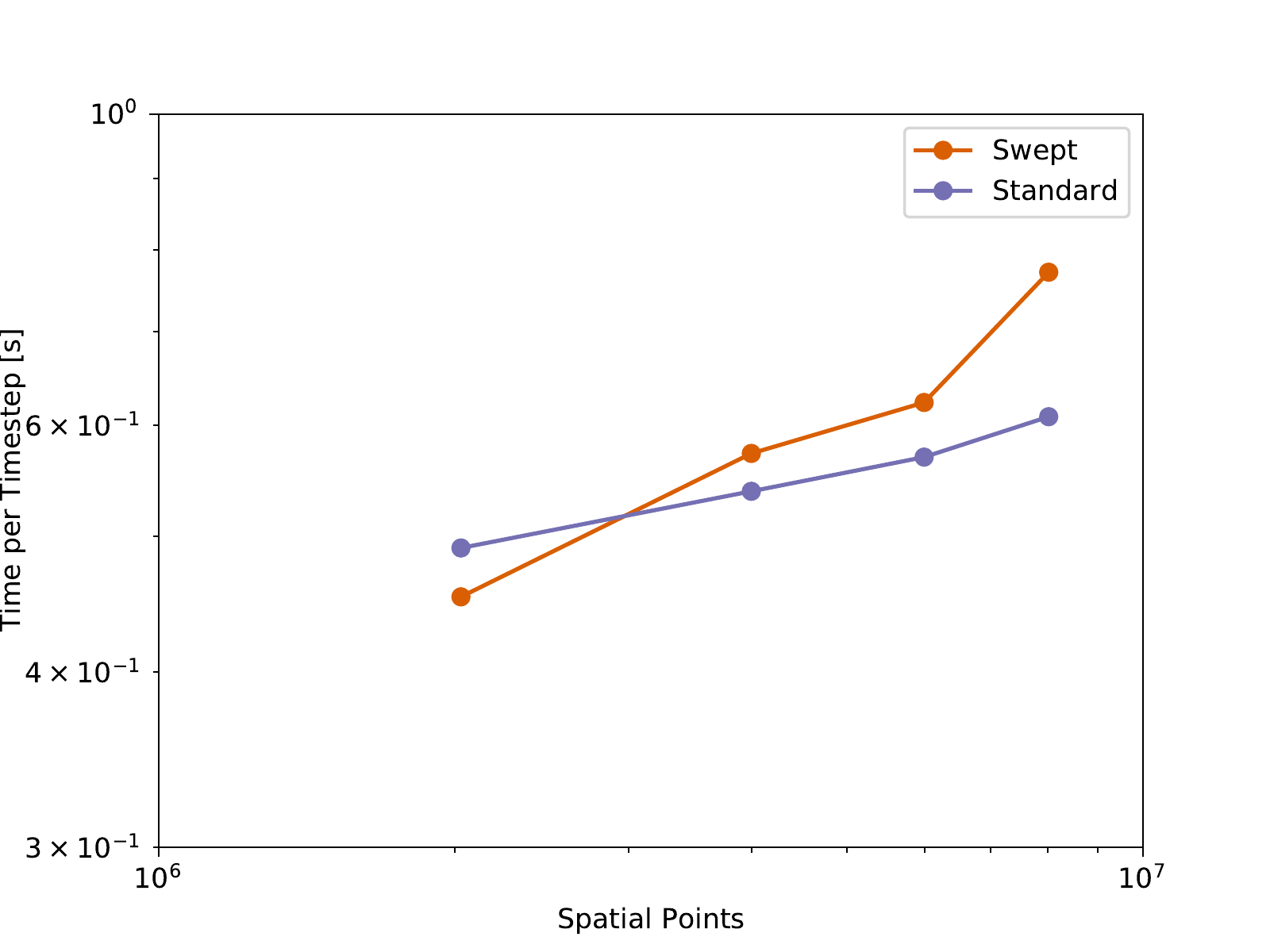}
    \caption{Weak scalability of the \Swept{} and \Standard{} algorithms for the heat equations.}
    \label{fig:scalabilityHeat}
\end{figure}

\begin{figure}[htbp]
    \centering
    \includegraphics[height=9cm,width=0.78\textwidth, trim={0cm 0.75cm 1cm 1.25cm},clip]{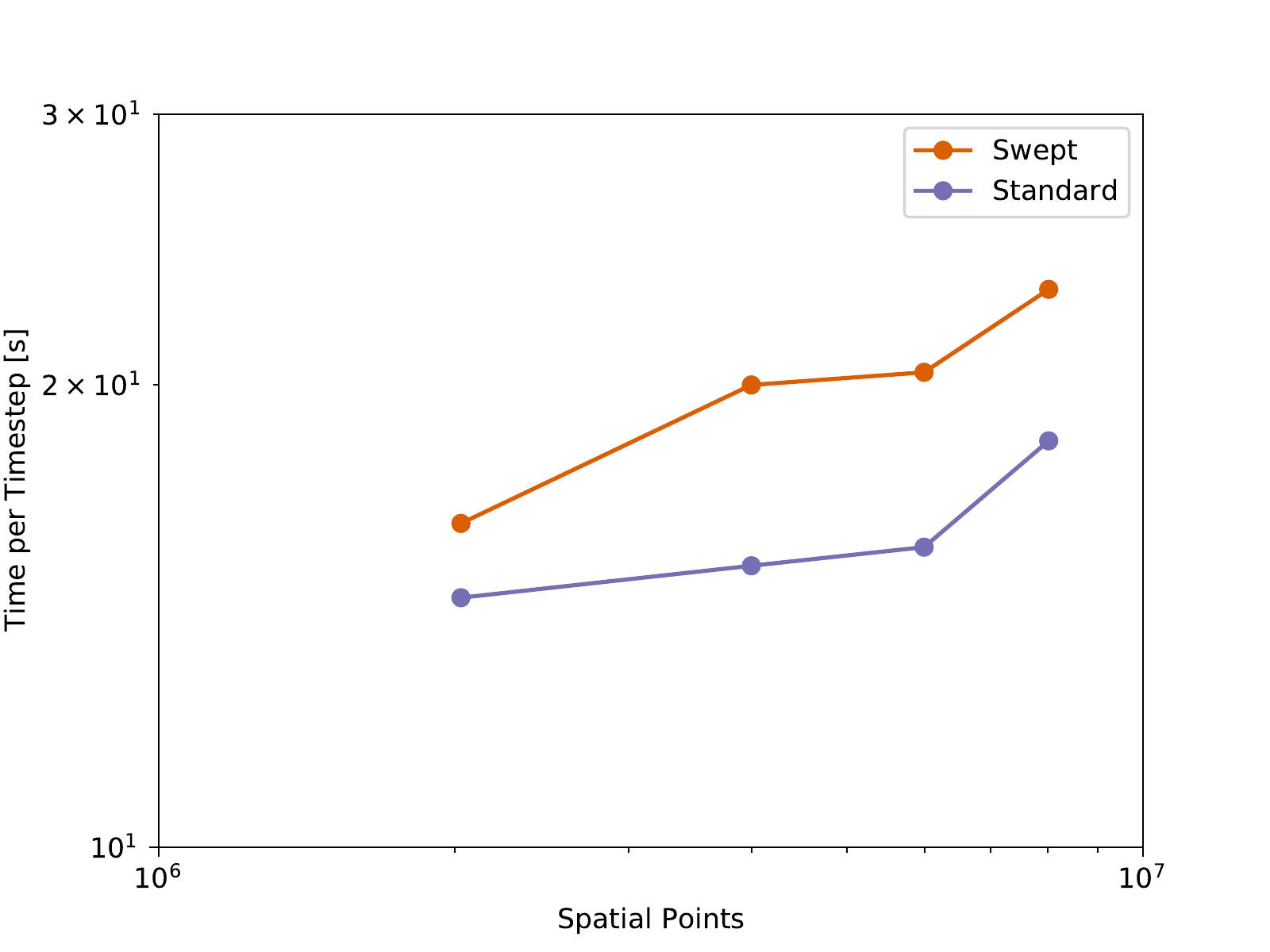}
    \caption{Weak scalability of the \Swept{} and \Standard{} algorithms for the Euler equations.}
    \label{fig:scalabilityEuler}
\end{figure}

These results raised some questions about the performance, e.g., why the swept rule performance decreases with increasing array size. We considered the weak scalability of the algorithms to explore this. Figures~\ref{fig:scalabilityHeat} and~\ref{fig:scalabilityEuler} show the time per timestep of each solver for the two problems. 
The \Standard{} solver shows better weak scalability in both cases. This suggests that something other than latency dominates the simulations so latency improvements are less noticeable.

\section{Discussion}
\label{discussion-section}

In regards to the heat diffusion equation, we see an overall range of speedup from 0.22--2.71 for the first hardware set and 0.79--1.32 for the second. However, the limits of speedup are outliers. These speedups are lower in comparison to the one-dimensional heterogeneous version which reported 1.9 to 23 times speedup \cite{Magee2020ApplyingSystems} and, the one-dimensional GPU version which reported 2 to 9 times speedup for similar applications \cite{Magee2018AcceleratingDecomposition}. 

For the Euler equations, we see an overall speedup range of 0.52--1.46 for the first set of hardware and 0.36--1.42 for the second. 
These speedups are better aligned with values reported in other studies. 
The two-dimensional CPU version reported a speedup of at most 4~\cite{Alhubail2018ThePDEs}, the one-dimensional GPU version reported 0.53--0.83~\cite{Magee2018AcceleratingDecomposition}, and the one-dimensional heterogeneous version reported 1.1--2.0 \cite{Magee2020ApplyingSystems}.

It is consistent between studies that the swept rule performance declines as numerical complexity increases. We suspect that the magnitude of speedups differ noticeably because performance is partially implementation based. Regardless, the swept rule can be helpful or harmful to simulation performance but testing is required to in the process. It depends on the problem solved, desired numerical scheme, and implementation.  
We see from the contours that the more complicated problem generally performs worse. In contrast, the heat equation shows quite a few areas of speed up for lower array sizes. 

The scalability results also help us to understand the swept rule's performance.
Figure~\ref{fig:scalabilityHeat} and~\ref{fig:scalabilityEuler} shows that generally the swept algorithm scales worse for both problems. Worse scalability is likely caused by bandwidth dominating the simulation. 
We suspect that the implementation of the communication step of the \Swept{} solver is responsible for this because it moves around more data than \Standard{}. 

The decline in performance with increasing in array sizes and scalability also supports our suspicion.
More data moves with larger array sizes, leading to higher bandwidth costs. In regards to scalability, notice how the \Swept{} scalability curve in Figure~\ref{fig:scalabilityEuler} is separated completely from the \Standard{} curve but this does not occur in Figure~\ref{fig:scalabilityHeat}. 
The Euler equations have roughly four times the amount of data to communicate making it more expensive. 

Other causes of poor performance could be steps that the \Swept{} solver takes but the \Standard{} does not. \pysweep{} determines and stores the appropriate blocks for communication prior to the simulation to avoid conditionals with the intention of boosting GPU performance. This process was parallelized as much as possible but there are some serial parts which undoubtedly reduce the performance of the algorithm. 

Scalability and array size aside, performance is also tied to the block size. The best case frequently occurs between intermediate block sizes because these cases balance thread divergence and latency improvements. The GPU portion naively removes points with conditional statements which causes some threads to diverge. 
So, while this achieves savings in latency, performance takes a hit due to thread divergence on GPUs.

\begin{figure}[htbp]
    \centering
    \includegraphics[height=9cm,width=0.78\textwidth, trim={0.5cm 0.4cm 0.5cm 0.2cm},clip]{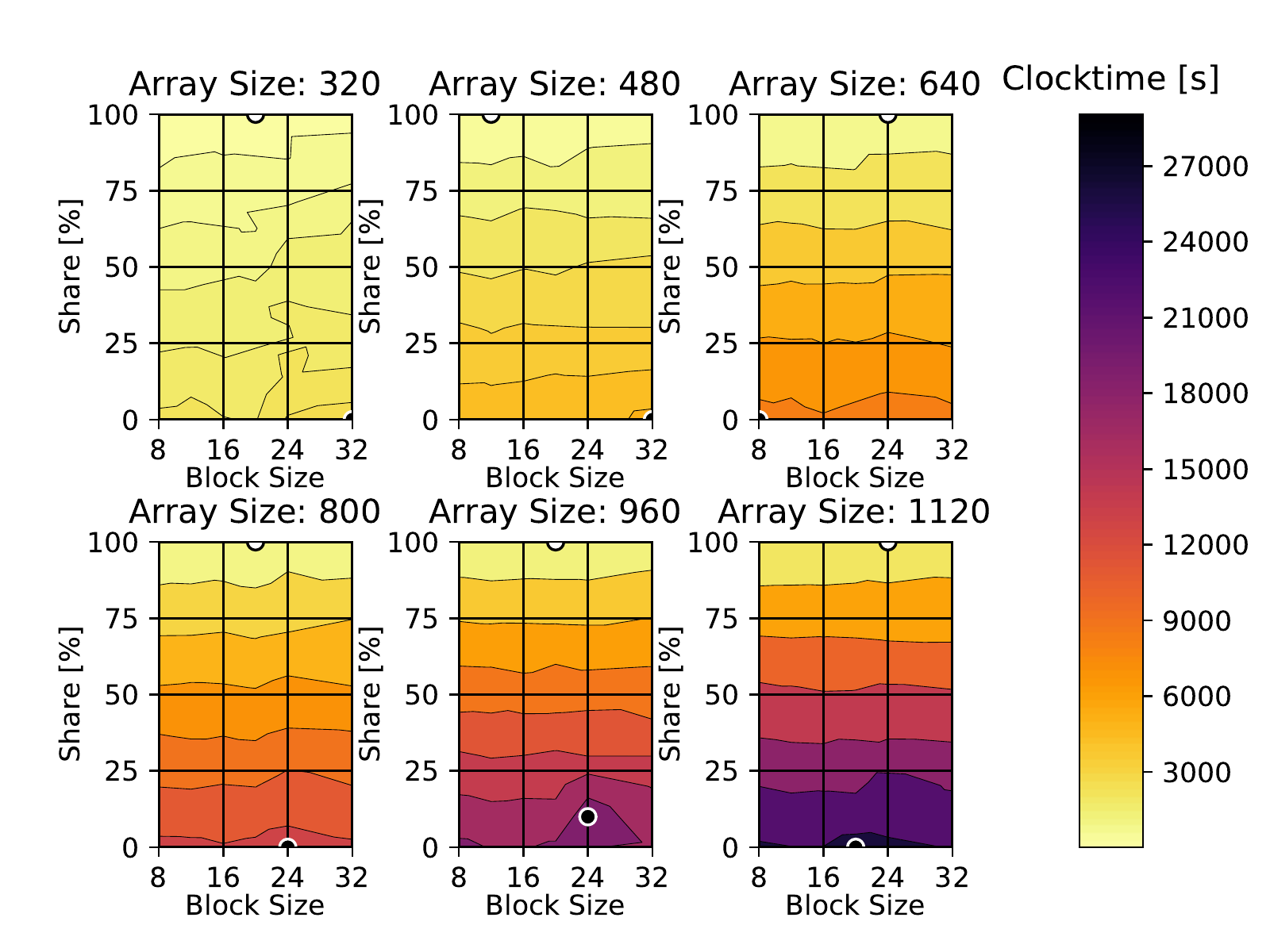}
    \caption{Clock time results  for the compressible Euler equations with \oldGPU{} GPUs and \oldCPU{} CPUs.}
    \label{fig:clocktimeOldEuler}
\end{figure}

The fastest option is typically the most desired outcome when it comes to high performance computing applications. While in some cases a GPU share of less than 100\% yield faster results, this is not one of those cases. Figure~\ref{fig:clocktimeOldEuler} demonstrates that the clock time only increases as share decreases. Note, figures for clock time in other cases were not presented because they show the same result: 100\% leads to the lowest clock time. 
However, the optimal configuration is useful if simulation requires data greater than the limit of the GPU(s).

\section{Conclusions}
\label{conclusions-section}

In this study we examined the performance of a two-dimensional swept rule solver on heterogeneous architectures. 
We tested our implementation over a range of GPU work shares, block sizes, and array sizes with two hardware configurations each containing two nodes. 
Our goal was to understand how the method performs based on these configurations.

Swept rule performance depends on the problem and scheme when using heterogeneous architecture. 
If possible, using solely GPUs provides the fastest run times for all cases we tested, and the swept rule can offer speedup in some of these pure-GPU cases. 
If GPUs cannot solely be used, speedup can still be obtained with optimal tuning, which depends on the input parameters. 
Generally, block sizes between 16--24 and a GPU share above 75\% are a good place to start.

Next, we found that hardware can affect swept-rule performance. 
The swept rule does show differing performance characteristics between different hardware combinations, but the major trends hold. 
However, the precise combination of performance parameters leading to optimal performance does vary between the hardware sets, and this should be considered when tuning for a given a simulation. Hardware-based performance differences also varied by the problem.

Overall, we conclude from this study that the performance of two-dimensional swept rule depends on the implementation, problem, and computing hardware. 
Speedup can be achieved but care should be taken when designing and implementing a solver. 
Finally, any practical use of the swept rule requires tuning input parameters to achieve optimal performance, 
and in some cases speedup at all over a traditional decomposition.

\vspace{6pt} 



\funding{This material is based upon work supported by NASA under award No. NNX15AU66A under the technical monitoring of Drs. Eric Nielsen and Mujeeb Malik.}

\dataavailability{All code to reproduce the data can be found at \github{}}.

\acknowledgments{We gratefully acknowledge the support of NVIDIA Corporation, who donated a Tesla K40c GPU used in developing this research.}

\appendixtitles{yes} 
\appendixstart
\appendix
\section{Heat Diffusion Equation}
\label{Heat-Diffusion}
The heat diffusion problem is as follows:
\begin{align*}
    &\rho c_p \frac{\partial T}{\partial t} = k\frac{\partial^2 T}{\partial x^2}+k\frac{\partial^2 T}{\partial y^2},\quad 0<x<1,\quad 0<y<1,\quad t>0\\
    &T(0,y,t) = T(1,y,t),\quad 0<x<1,\quad t>0\\
    &T(x,0,t) = T(x,1,t),\quad 0<y<1,\quad t>0\\
    &T(x,y,0) = \sin(2\pi x) \sin(2\pi y) \;, \quad 0\leq x\leq 1,\quad 0 \leq y \leq 1
\end{align*}
with an analytical solution of
\begin{equation*}
\label{heat-analyt}
    T(x,y,t) = \sin(2\pi x) \sin(2\pi y) e^{-8\pi^2\alpha t}
\end{equation*}
and an update equation of 
\begin{equation*}
    \label{heat-update}
    T_{i,j}^{k+1} = T_{i,j}^{k}+\frac{\alpha \Delta t}{\Delta x^2}\big(T_{i+1,j}^{k}-2T_{i,j}^{k}+T_{i-1,j}^{k}\big)+\frac{\alpha \Delta t}{\Delta y^2}\big(T_{i,j+1}^{k}-2T_{i,j}^{k}+T_{i,j-1}^{k}\big) \;.
\end{equation*}

\section{Compressible Euler Vortex}
\label{Compressible-Euler}
The compressible Euler equations and the equation of state used are as follows where subscripts represent derivatives with respect to the spatial dimensions ($x$ and $y$) or time ($t$):
\begin{align*}
    &\rho_t  = (\rho u)_x + (\rho v)_y \\ 
    &(\rho u)_t  = (\rho u+P)_x + (\rho u v)_y\\
    &(\rho v)_t  = (\rho v+P)_y + (\rho u v)_x\\
    &E_t = ((E+P)u)_x+((E+P)v)_y\\
    &E = \frac{P}{\gamma -1}+\frac{1}{2}\rho(u^2+v^2)
\end{align*}
The analytical solution to the isentropic Euler vortex was used as the initial conditions and in verification. The analytical solution was developed from Spiegel et al. \cite{SpiegelAMethods}. The solution is simple in the sense that it is just translation of the initial conditions. It involves superimposing perturbations in the form
\begin{align*}
    &\delta u = -\frac{y}{R}\Omega\\
    &\delta v = -\frac{x}{R}\Omega\\
    &\delta T = -\frac{\gamma-1}{2} \Omega^2\\
    &\Omega = \beta e^{f}\text{, where } f(x,y) = -\frac{1}{2\sigma^2}\Bigg[\bigg(\frac{x}{R}^2\bigg)+\frac{y}{R}^2\bigg)\Bigg].
\end{align*}
The initial conditions are then 
\begin{align*}
    &\rho_0 = (1+\delta T)^{\frac{1}{\gamma-1}},\\
    &u_0 = M_\infty \cos(\alpha)+\delta u,\\
    &v_0 = M_\infty \sin(\alpha)+\delta v,\\
    &p_0 = \frac{1}{\gamma}(1+\delta T)^\frac{\gamma}{\gamma-1} ;,
\end{align*}
where Table~\ref{tab:eulerVortex} shows the specific values used.

\begin{table}[htb]
    \centering
    \begin{tabular}{@{}ccccccccc@{}}
    \toprule
    $\alpha$ & $M_\infty$  & $\rho_\infty$ & $p_\infty$ & $T_\infty$ & $R$ & $\sigma$ & $\beta$ & $L$ \\ 
    \midrule
    $45^\circ$ &  $\sqrt{\frac{2}{\gamma}}$ & $1$ & $1$ & $1$ & $1$ & $1$ & $M_\infty \frac{5\sqrt{2}}{4\pi}e^{1/2}$ & $5$\\
    \bottomrule
    \end{tabular}
    \caption{Conditions used in analytical solution \cite{shu1998essentially}}
    \label{tab:eulerVortex}
\end{table}

Similar to the heat diffusion equation, periodic boundary conditions were implemented. 
We implemented a similar numerical scheme to that of Magee et al.~\cite{Magee2018AcceleratingDecomposition}, except extended it into two dimensions. 
A five-point finite volume method was used in space with a minmod flux limiter and second-order Runge--Kutta scheme was used in time. 
Consider equations of the form
\begin{equation*}
    \frac{\partial Q}{\partial t}+\frac{\partial F}{\partial x}+\frac{\partial G}{\partial y} = 0 \;,
\end{equation*}
where
\begin{equation*}
    Q = \begin{bmatrix}
        \rho\\
        \rho u\\
        \rho v\\
        E
        \end{bmatrix}
        \text{, }
    F = \begin{bmatrix}
    \rho u\\
    \rho u^2+p\\
    \rho u v\\
    (E+p)u
    \end{bmatrix}
    \text{, and }
    G = \begin{bmatrix}
    \rho v\\
    \rho u v\\
    \rho v^2+p\\
    (E+p)v
    \end{bmatrix} \;.
\end{equation*}
The minmod limiter uses the pressure ratio to compute reconstructed values on the cell boundaries:
\begin{align*}
   & P_{r,i} = \frac{P_{i+1}-P_{i}}{P_{i}-P_{i-1}},\\
    &Q_n^{i} = \begin{cases} 
      Q_o^{i}+\frac{\min(P_{r,i}^1,1)}{2}(Q_o^{i+1)}-Q_o^{i}) & 0 < P_{r,i} < \infty \\
      Q_0^{i} \\
   \end{cases},\\
    &Q_n^{i+1} = \begin{cases} 
      Q_o^{i+1}+\frac{\min(P_{r,i+1}^{-1},1)}{2}(Q_o^{i}-Q_o^{i+1}) & 0 < P_{r,i}^{-1} < \infty \\
      Q_0^{i+1} \\
   \end{cases} \;.
\end{align*}
The flux is then calculated with the reconstructed values for $i$ and $i+1$ accordingly and used to step in time:
\begin{align*}
    &F_{i+1} = \frac{1}{2}\big( F(Q^{i+1})+F(Q^{i})+r_{x,sp}(Q^{i}-Q^{i+1})\big)\\
    &G_{i+1} = \frac{1}{2}\big( G(Q^{i+1})+G(Q^{i})+r_{y,sp}(Q^{i}-Q^{i+1})\big),\\
\end{align*}
where $r_{i,sp}$ is the spectral radius or largest eigenvalue for the appropriate Jacobian matrix. 
These flux calculations are then used to apply the second-order Runge--Kutta in time:
\begin{align*}
    &Q_i^{*} = Q_i^{n}+\frac{\Delta t}{2\Delta x}(F_{i+1/2}(Q^n)+F_{i-1/2}(Q^n))+\frac{\Delta t}{2\Delta y}(G_{i+1/2}(Q^n)+G_{i-1/2}(Q^n)),\\
    &Q_i^{n+1} = Q_i^{n}+\frac{\Delta t}{\Delta x}(F_{i+1/2}(Q^*)+F_{i-1/2}(Q^*))+\frac{\Delta t}{\Delta y}(G_{i+1/2}(Q^*)+G_{i-1/2}(Q^*)) \;.
\end{align*}



\end{paracol}
\reftitle{References}


\externalbibliography{yes}
\bibliography{references}

%



\end{document}